\documentclass[aps,pra,twocolumn,showpacs]{revtex4}

\usepackage[dvips]{graphicx}

\input{epsf}

\begin{document}

\title{Quantum Monte Carlo simulation of a two-dimensional Bose gas}

\author{S. Pilati$^{(1,2)}$, J. Boronat$^{(2)}$, J. Casulleras$^{(2)}$, and S. Giorgini$^{(3,1)}$}
\address{$^{(1)}$Dipartimento di Fisica, Universit\`a di Trento and CRS-BEC INFM, I-38050 Povo, Italy\\
$^{(2)}$Departament de F\'{\i}sica i Enginyeria Nuclear, Campus Nord B4-B5, Universitat Polit\`ecnica 
de Catalunya, E-08034 Barcelona, Spain\\
$^{(3)}$ JILA, University of Colorado, Boulder, CO 80309-0440, U.S.A.}

\date{\today}

\begin{abstract}
The equation of state of a homogeneous two-dimensional Bose gas is calculated using quantum Monte Carlo 
methods. The low-density universal behavior is investigated using different interatomic
model potentials, both finite-ranged and strictly repulsive and zero-ranged supporting a bound state. The condensate 
fraction and the pair distribution function are calculated as a function of the gas parameter, ranging from the 
dilute to the strongly correlated regime. In the case of the zero-range pseudopotential we discuss the stability 
of the gas-like state for large values of the two-dimensional scattering length, and we calculate the critical 
density where the system becomes unstable against cluster formation.     
\end{abstract}


\maketitle

\section{Introduction}
\label{Introduction}
Degenerate low-dimensional gases are presently attracting considerable interest as model systems to investigate 
beyond mean-field effects and phenomena where many-body correlations, thermal and quantum fluctuations play a relevant 
role~\cite{LesHouches}. In two dimensions (2D), it is well known that thermal excitations destroy long-range order 
and Bose-Einstein condensation (BEC) can not exist in large bosonic systems at finite temperature~\cite{Hohenberg}. 
Nevertheless, a defect-mediated phase transition from a high-temperature normal fluid to a low-temperature superfluid 
was predicted by Berezinskii, Kosterlitz and Thouless (BKT)~\cite{BKT} and has been observed in thin films of liquid 
$^4$He~\cite{Reppy}. At zero temperature, fluctuations are instead negligible and BEC is possible. The ground-state 
energy per particle of a homogeneous dilute Bose gas in 2D has been first calculated by Schick~\cite{Schick} and is 
given by
\begin{equation}
\frac{E_{MF}}{N}=\frac{2\pi\hbar^2}{m}\frac{n}{\log(1/na_{2D}^2)} \;,
\label{eMF}
\end{equation}    
where $m$ is  the mass of the particles, $n$ is the number density and $a_{2D}$ is the 2D scattering length. The above 
result holds for small values of the gas parameter,  $na_{2D}^2\ll 1$, and can be obtained within a mean-field 
approximation using the coupling constant
\begin{equation}
g_{2D}=\frac{4\pi\hbar^2}{m}\frac{1}{\log(1/na_{2D}^2)} \;.
\label{g2D}
\end{equation}
The presence at low energy of a density dependent coupling constant is a peculiar feature of 2D systems. 

Bose gases in quasi-2D have been realized with spin-polarized hydrogen on a liquid-helium surface~\cite{Safonov} and 
with alkali-metal atoms confined in highly anisotropic harmonic traps~\cite{EXP,Grimm,Foot}. In these systems, the transverse 
motion of the atoms is frozen to zero-point oscillations resulting in fully 2D kinematics, but the interatomic 
scattering processes are still three-dimensional (3D). In fact, the 3D s-wave scattering length $a_{3D}$ is always much smaller 
than the characteristic length of the tight transverse confinement. Theoretical studies of quasi-2D trapped systems predict 
the existence of a low-temperature quasicondensate phase~\cite{Popov,Petrov1} whose properties have been investigated 
using mean-field approaches~\cite{MF}. The nature of the  transition in confined systems and  whether it belongs to 
the  BEC or BKT universality class is still an open problem both experimentally and theoretically.

An important theoretical prediction for quasi-2D Bose gases in harmonic traps is the renormalization of the coupling
constant due to the tight confinement in one direction. To logarithmic accuracy the result is given 
by~\cite{Petrov1,Petrov2}
\begin{equation}
g_{2D}=\frac{2\sqrt{2\pi}\hbar^2}{m}\frac{1}{\frac{a_z}{a_{3D}}+\frac{1}{\sqrt{2\pi}}\log(1/n(0)a_z^2)} \;,
\label{g2D1}
\end{equation}
where $a_z=\sqrt{\hbar/m\omega_z}$ is the harmonic oscillator length, fixed by the frequency $\omega_z$ of the tight 
transverse confinement, and $n(0)$ is the 2D density in the center of the trap. If $a_{3D}\ll a_z$, one can neglect the 
$\log$ term in Eq. (\ref{g2D1}) and the resulting coupling constant is determined by the 3D scattering length. On the 
contrary, if $a_{3D}\gg a_z$, $g_{2D}$ becomes independent of the value of $a_{3D}$ and a regime of  
pure 2D scattering is achieved with $a_{2D}=a_z$. This universal regime, where the properties of the system only 
depend on density and not on interaction, is highly interesting and can be realized in trapped gases by using a 
Feshbach resonance~\cite{Feshbach} to achieve large values for $a_{3D}$.

In this work we investigate the ground-state properties of a strictly 2D homogeneous Bose gas using quantum Monte Carlo
techniques. We determine the equation of state over a wide range of values of the gas parameter from the extremely dilute
regime, $na_{2D}^2\sim 10^{-7}$, to the strongly correlated regime, $na_{2D}^2\sim 0.1$. The calculations of the energy
per particle are  carried out using three different interatomic model potentials: two repulsive finite-ranged potentials 
and a zero-range pseudopotential which supports a two-body bound state. We investigate beyond mean-field corrections to 
the equation of state (\ref{eMF}) and their dependence on the gas parameter $na_{2D}^2$. In the 
case of the zero-range pseudopotential, from a study of the system compressibility as a function of the gas parameter we 
estimate that $na_{2D}^2\simeq 0.04$ is the critical density at which the gas-like state becomes unstable against cluster 
formation. This result puts an upper limit to the value of density that can be reached in harmonic traps when one enters 
the regime of pure 2D scattering. We also report results of the pair distribution function and the condensate fraction as 
a function of the gas parameter.

The structure of the paper is as follows. In Section~\ref{Method} we introduce the many-body Hamiltonian and the model 
interatomic potentials we use in the simulations. We also discuss the quantum Monte Carlo methods employed in the present 
study and the choice of the trial wave functions. Section~\ref{Results} containes the results obtained and 
Section~\ref{Conclusions} draws our conclusions.

\section{Method}
\label{Method}
We consider a homogeneous system of $N$ spinless bosons in 2D described by the many-body Hamiltonian
\begin{equation}
H=-\frac{\hbar^2}{2m}\sum_{i=1}^{N}\nabla_i^2+\sum_{i<j}V({\bf r}_i-{\bf r}_j)\;,
\label{ham}
\end{equation}
where ${\bf r}_i=x_i\hat{\bf i}+y_i\hat{\bf j}$ denotes the 2D position vector of the $i$-th particle and $V({\bf r})$ is 
the two-body interatomic potential. We use different choices for the interatomic potential $V({\bf r})$:

(i) Hard-disk (HD) potential defined by
\begin{eqnarray}
V^{HD}({\bf r}) = \left\{ 
\begin{array}{cll} \infty           & (r<a_{2D}) \\
                      0             & (r>a_{2D}) \;,
\end{array}\right.
\label{hdpot}
\end{eqnarray}
in which case the hard-disk diameter corresponds to the scattering length $a_{2D}$. 

(ii) Soft-disk (SD) potential defined by
\begin{eqnarray}
V^{SD}({\bf r}) = \left\{ 
\begin{array}{cll}   V_0            & (r<R) \\
                      0             & (r>R) \;,
\end{array}\right.
\label{sdpot}
\end{eqnarray}
in which case the scattering length is given in terms of the modified Bessel function $I_0(x)$ and its derivative 
$I_0^\prime(x)$,
\begin{equation}
a_{2D}=R \exp\left[-\frac{1}{RK_0}\frac{I_0(RK_0)}{I_0^\prime(RK_0)}\right] \;,
\label{sda2D}
\end{equation}
where $K_0^2=mV_0/\hbar^2$ with $V_0>0$.
For a finite $V_0$ one has always $R>a_{2D}$; if $V_0\to +\infty$ the SD and HD potentials coincide with $a_{2D}=R$. 
In the present calculations, the 
range of the SD potential is kept fixed to the value $R=5a_{2D}$ and the height $V_0$ is determined through Eq. (\ref{sda2D}) 
to give the desired value of $a_{2D}$.

(iii) Zero-range pseudopotential (PP) defined by~\cite{OLP}
\begin{equation}
V^{PP}({\bf r}) = -\frac{2\pi\hbar^2}{m}\frac{\delta({\bf r})}{\log(qa_{2D})}\left[1-\log(qr)r\frac{\partial}
{\partial r}\right] \;, 
\label{pppot}
\end{equation} 
where the second term in the square brackets is a regularizing operator with $q$ an arbitrary wave vector. The potential 
$V^{PP}(r)$ supports a two-body bound state with energy $\epsilon_b=-4\hbar^2/(ma_{2D}^2e^{2\gamma})$ and wave function 
$f_b(r)=K_0(2r/(a_{2D}e^\gamma))$, where $K_0(x)$ is the modified Bessel function and $\gamma\simeq 0.577$ is the Euler 
constant. 

The calculations with the repulsive potentials $V^{HD}(r)$ and $V^{SD}(r)$ are carried out using the diffusion Monte 
Carlo (DMC) method. This technique solves the many-body time-independent Schr\"odinger equation by evolving the 
function $f({\bf R},\tau)=\psi_T({\bf R})\Psi({\bf R},\tau)$ in imaginary time $\tau=it/\hbar$ according to the 
time-dependent Schr\"odinger equation
\begin{eqnarray}
-\frac{\partial f({\bf R},\tau)}{\partial\tau}= &-& D\nabla_{\bf R}^2 f({\bf R},\tau) + D \nabla_{\bf R}[{\bf F}({\bf R})
f({\bf R},\tau)] \nonumber \\
&+& [E_L({\bf R})-E_{ref}]f({\bf R},\tau) \;.
\label{FNDMC}
\end{eqnarray}
Here $\Psi({\bf R},\tau)$ denotes the wave function of the system and $\psi_T({\bf R})$ is a trial function used for 
importance sampling. In the above equation ${\bf R}=({\bf r}_1,...,{\bf r}_N)$, $E_L({\bf R})=
\psi_T({\bf R})^{-1}H\psi_T({\bf R})$ denotes the local energy, ${\bf F}({\bf R})=2\psi_T({\bf R})^{-1}\nabla_{\bf R}
\psi_T({\bf R})$ is the quantum drift force, $D=\hbar^2/(2m)$ plays the role of an effective diffusion constant, and 
$E_{ref}$ is a reference energy introduced to stabilize the numerics. The energy and other observables of the ground state 
of the system are calculated from averages over the asymptotic distribution function $f({\bf R},\tau\to\infty)$. 
Apart from statistical errors, the DMC method determines the ground-state energy of a system of $N$ 
bosons exactly.

In the case of the pseudopotential $V^{PP}(r)$, Eq. (\ref{pppot}), the gas-like state is not the ground state of the
system, but is a highly-excited state of the Hamiltonian. We expect, however, that for small values of the gas 
parameter, $na_{2D}^2\ll 1$, the gas-like state is stable against formation of many-body bound states.
The calculations with the PP potential are performed at the level of the variational Monte Carlo (VMC) method to avoid 
the technical difficulties that arise in the study of excited states in the DMC method. Nevertheless, as it will be shown 
in Section~\ref{Results}, the accuracy achieved by using this variational approach is very high. The VMC technique is 
based on the variational principle
\begin{equation}
\frac{\langle\psi_T|H|\psi_T\rangle}{\langle\psi_T|\psi_T\rangle}\geq E\;,
\label{VMC}
\end{equation} 
which is here applied to excited states of the Hamiltonian $H$. For a trial wave function $\psi_T$ with a given symmetry, 
the variational 
estimate provides an upper bound to the energy $E$ of the lowest excited state of the Hamiltonian $H$ with that symmetry.

In all the calculations the trial wave function $\psi_T({\bf R})$ is taken of the Jastrow form
\begin{equation}
\psi_T({\bf r}_1,...,{\bf r}_N)=\prod_{i<j}f(r_{ij})\;,
\label{trialwf}
\end{equation}
where the correlation factor $f(r)$ is chosen as the solution of the two-body Schr\"odinger equation subject to the 
boundary conditions $f(r\ge d)=1$ and $f^\prime(r\ge d)=0$ on the wave 
function and its first derivative respectively, with $d$ a variational parameter. For the repulsive potentials, HD and SD,
$f(r)$ corresponds to the ground-state two-body wave function. On the contrary, in the case of the PP interaction, $f(r)$
is chosen as the two-body wave function corresponding to the lowest excited state with positive energy. The pseudopotential 
$V^{PP}$ imposes the following boundary condition on the pair wave function for zero interparticle distance
\begin{equation}
\frac{\left[rf^\prime(r)\right]_{r=0}}{\left[f(r)-\log(qr)rf^\prime(r)\right]_{r=0}}=-\frac{1}{\log(qa_{2D})}\;.
\label{boundary}
\end{equation}
The free two-particle problem in 2D $-\hbar^2/m\nabla^2f(r)=\hbar^2k^2/mf(r)$, holding for $r\neq 0$, gives the 
general solution
\begin{equation}
f(r)=AJ_0(kr)+BY_0(kr) \;,
\label{freesol}
\end{equation}
in terms of the $J_0(x)$ and $Y_0(x)$ Bessel functions.
The boundary condition (\ref{boundary}) is imposed by the relation
\begin{equation}
a_{2D}=\frac{2e^{-\gamma}}{k}\exp\left(-\frac{\pi A}{2B}\right) \;.
\label{boundary1}
\end{equation}
In addition we require that $f(r)$ possess only one node for distances  $0<r<d$. The boundary conditions at $r=d$ 
together with (\ref{boundary1}) fix the values of the constants $A$ and $B$ and of the wavevector $k$. 
If $d\gg a_{2D}$, the eigenenergy $\hbar^2k^2/m$ is small and the node of $f(r)$ is located at $r=a_{2D}$. 

All the Monte Carlo calculations have been carried out considering periodic boundary conditions with a cubic simulation box 
whose size $L$ is fixed by the density $n$ and by the total number of atoms: $n=N/L^2$. The healing length $d$ is considered
everywhere $d=L/2$ since a VMC optimization has shown that this is a good option.

\begin{table}
\centering
\begin{ruledtabular}
\caption{Energy per particle $E/N$ corresponding to the model potentials $V^{HD}(r)$ (DMC results), $V^{SD}(r)$ (DMC results) 
and $V^{PP}(r)$ (VMC results) [energies are in units of $\hbar^2/(2ma_{2D}^2)$].}
\begin{tabular}{cccc}
$na_{2D}^2$ & $HD$ & $SD$ & $PP$ \\  
\hline
$1\times10^{-7}$   & $7.738(3)\times10^{-8}$ & $7.732(6)\times10^{-8}$ &           \\
$3\times10^{-7}$   & $2.500(1)\times10^{-7}$ & $2.498(1)\times10^{-7}$ &           \\
$1\times10^{-6}$   & $9.103(5)\times10^{-7}$ & $0.909(1)\times10^{-6}$ & $9.162(7)\times10^{-7}$ \\
$3\times10^{-6}$   & $2.982(2)\times10^{-6}$ & $2.977(4)\times10^{-6}$ &           \\
$1\times10^{-5}$   & $1.108(2)\times10^{-5}$ & $1.106(1)\times10^{-5}$ & $1.115(1)\times10^{-5}$ \\
$3\times10^{-5}$   & $3.710(4)\times10^{-5}$ & $3.702(2)\times10^{-5}$ &           \\
$1\times10^{-4}$   & $1.417(1)\times10^{-4}$ & $1.415(1)\times10^{-4}$ & $1.428(1)\times10^{-4}$ \\
$3\times10^{-4}$   & $4.922(5)\times10^{-4}$ & $4.910(2)\times10^{-4}$ &           \\
$1\times10^{-3}$   & $1.981(4)\times10^{-3}$ & $1.972(1)\times10^{-3}$ & $1.994(1)\times10^{-3}$ \\
$3\times10^{-3}$   & $7.34(2)\times10^{-3}$  & $7.224(3)\times10^{-3}$ &           \\
$1\times10^{-2}$   & $3.296(5)\times10^{-2}$ & $3.057(2)\times10^{-2}$ & $3.304(4)\times10^{-2}$ \\
$3\times10^{-2}$   & $1.436(2)\times10^{-1}$ & $1.105(1)\times10^{-1}$ & $1.369(6)\times10^{-1}$ \\
$1\times10^{-1}$   & $8.94(2)\times10^{-1}$  & $4.161(4)\times10^{-1}$ &           
\label{tab1}
\end{tabular}
\end{ruledtabular}
\end{table}

\begin{figure}
\begin{center}
\includegraphics*[width=7cm]{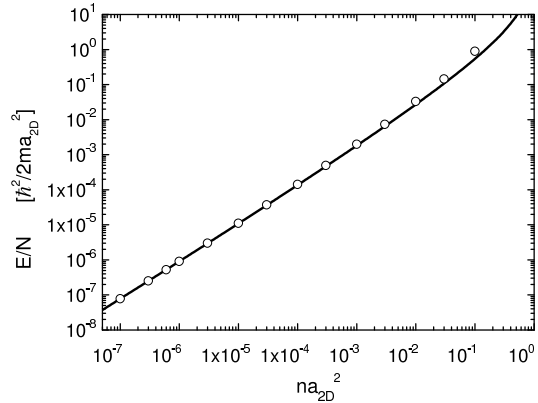}
\caption{Energy per particle as a function of the gas parameter for the HD potential. Solid line: mean-field result, Eq. 
(\ref{eMF}). Error bars are smaller than the size of the symbols.}
\label{fig1}
\end{center}
\end{figure}

\begin{figure}
\begin{center}
\includegraphics*[width=7cm]{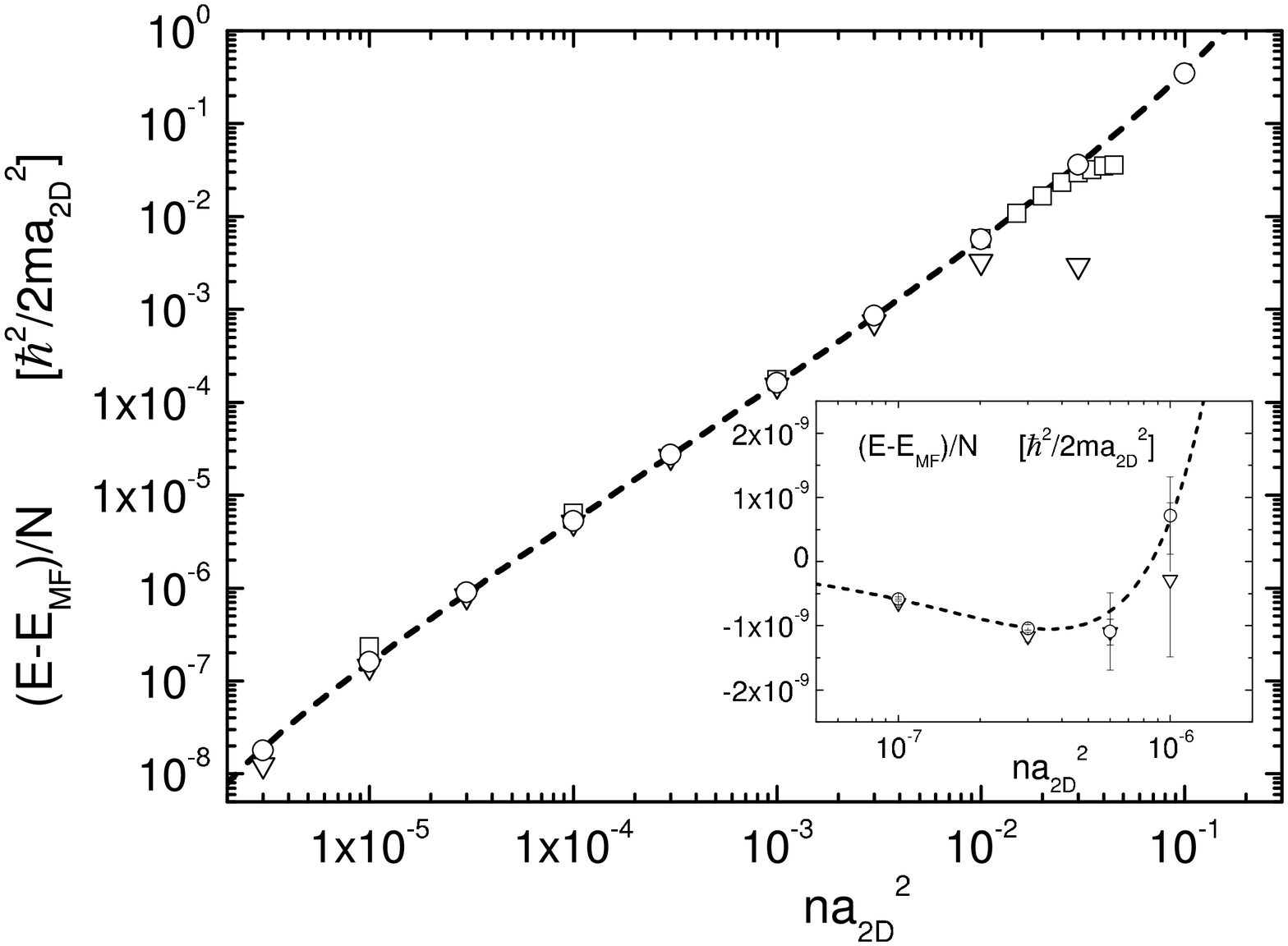}
\caption{Beyond mean-field corrections to the equation of state. Circles, HD potential; triangles: SD potential; squares: 
PP potential. Dashed line: numerical fit to the results of the HD potential, Eq. (\ref{FIT}). Error bars are smaller than 
the size of the symbols. Inset: region of extremely dilute systems $na_{2D}^2\le 10^{-6}$.}
\label{fig2}
\end{center}
\end{figure}

\begin{figure}
\begin{center}
\includegraphics*[width=7cm]{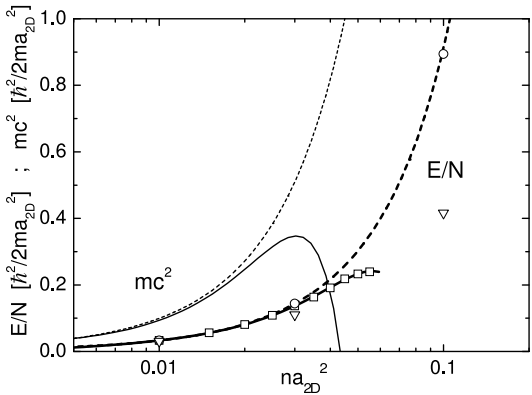}
\caption{Energy per particle and inverse compressibility as a function of the gas parameter. Symbols are as in 
Fig.~\ref{fig2}. Thick and thin dashed lines: best fit (\ref{FIT}) to the HD equation of state and corresponding $mc^2$, 
respectively. Thick and thin solid line: polynomial best fit to the PP equation of state and corresponding $mc^2$, 
respectively. Error bars are smaller than the size of the symbols.}
\label{fig3}
\end{center}
\end{figure}

\begin{figure}
\begin{center}
\includegraphics*[width=7cm]{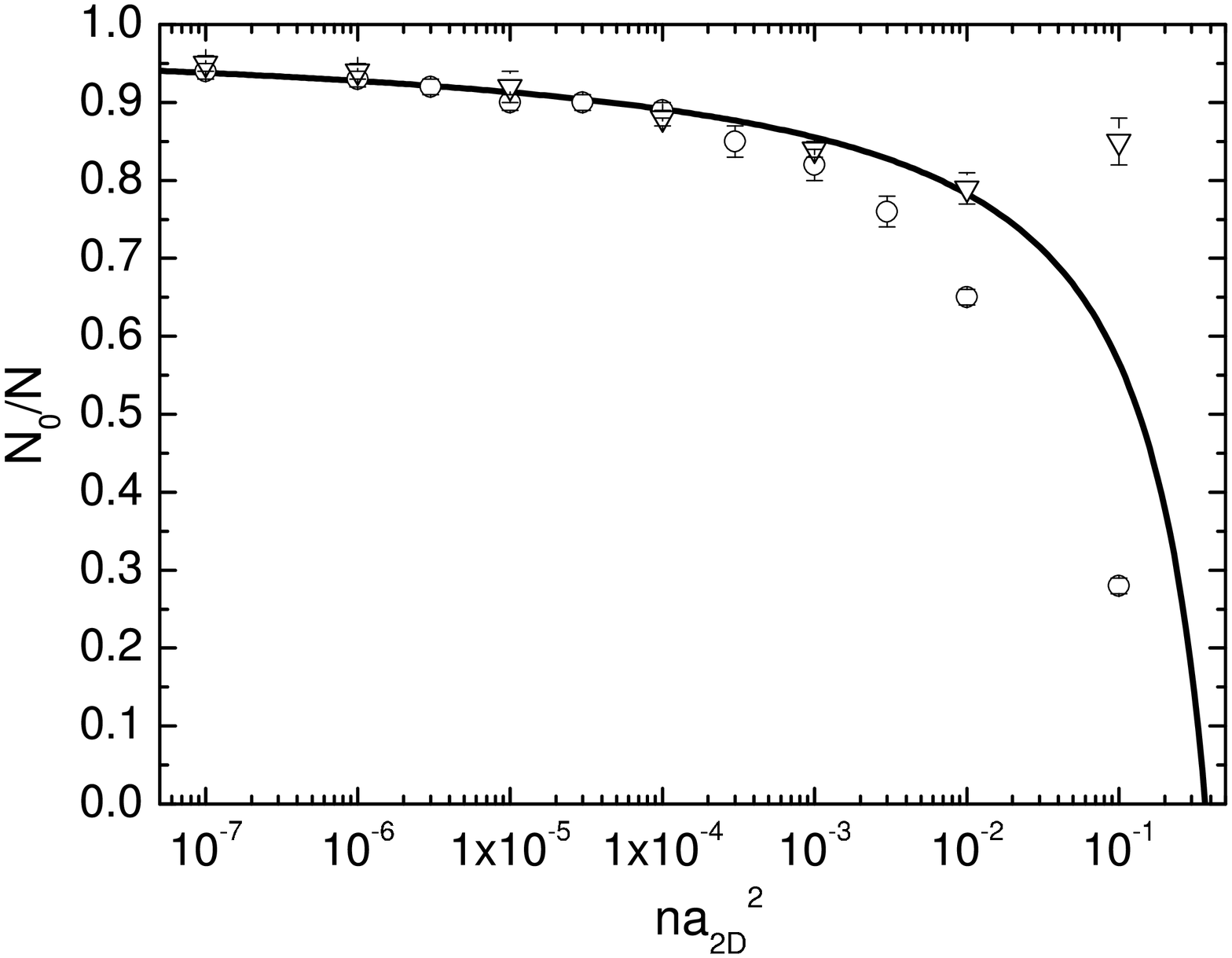}
\caption{Condensate fraction as a function of the gas parameter. Circles, HD potential; triangles, SD potential. Solid 
line: result from Bogoliubov theory, Eq. (\ref{CF}).}
\label{fig4}
\end{center}
\end{figure}

\begin{figure}
\begin{center}
\includegraphics*[width=7cm]{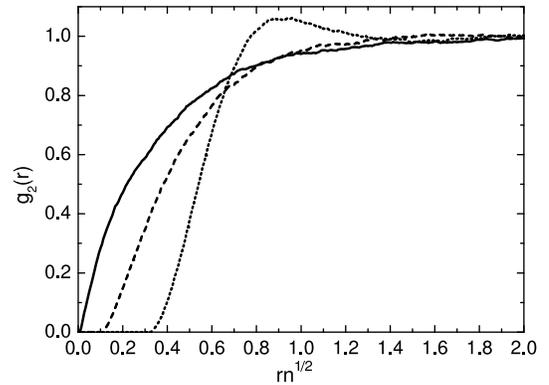}
\caption{Pair distribution function of a HD gas obtained for three values of the gas parameter: $na_{2D}^2=10^{-4}$, solid 
line; $na_{2D}^2=0.01$, dashed line; $na_{2D}^2=0.1$, dotted line.}
\label{fig5}
\end{center}
\end{figure}

\section{Results}
\label{Results} 
The numerical simulations are carried out with $N=512$ particles for the lowest densities ($na_{2D}^2\le 10^{-4}$) and 
with $N=256$ particles for the rest. 
The DMC results for the energy per particle corresponding to the HD potential are reported in Table~\ref{tab1} as a 
function of the gas parameter. Variational results on the same system have been obtained in the dilute regime by Mazzanti 
{\it et al.}~\cite{Mazzanti} using Correlated Basis Functions theory and at high densities by Lei Xing~\cite{MCARLO1} and 
Sol\'{\i}s~\cite{MCARLO2} using VMC. In Fig.~\ref{fig1}, the DMC results for the HD system are compared with the mean-field 
result, Eq. (\ref{eMF}). We find a remarkably 
good agreement up to large values of $na_{2D}^2$. It is worth noticing that the range of validity of the mean-field 
approximation is in 2D significantly larger than the one observed in 3D~\cite{US}.
Beyond mean-field corrections to the equation of state are 
shown in Fig.~\ref{fig2}, where we plot the difference $(E-E_{MF})/N$ for the three model potentials: $V^{HD}(r)$ 
(DMC results), $V^{SD}(r)$ (DMC results) and $V^{PP}(r)$ (VMC results). Focusing on the exact DMC results, one can notice the 
universal behavior of 
the equation of state for $na_{2D}^2\ll 1$. The results of the HD and SD model potentials are practically indistinguishable up 
to $na_{2D}^2\sim 10^{-2}$ (see Table~\ref{tab1}), showing that beyond mean-field corrections to Eq. (\ref{eMF}) only depend on 
the value of the gas parameter 
$na_{2D}^2$ and not on the details of the interatomic potential. By increasing $na_{2D}^2$ the results obtained with the SD 
potential (triangles in Fig.~\ref{fig2}) start to deviate from the ones corresponding to the HD potential (circles in 
Fig.~\ref{fig2}) approximately about $na_{2D}^2\sim 0.01$. The VMC results for the PP potential are upper bounds to the 
eigenenergy of the gas-like state. However, looking at Table~\ref{tab1} and Fig.~\ref{fig2}, one can see that they are nearly
indistinguishable from the exact HD and SD energies in the low-density regime. A similar accuracy for the energy of the gas-like
state is expected also for larger values of the gas parameter, where the results of the PP interaction start to separate from
the HD ones as the effects of the finite interaction range become more important. It is also worth 
noticing that for the smallest values of $na_{2D}^2$ (inset of Fig.~\ref{fig2}) the beyond mean-field correction is 
negative. This is consistent with the perturbation expansion~\cite{PEXP} 
\begin{equation}
\frac{E-E_{MF}}{N}=-\frac{2\pi\hbar^2n}{m}\frac{\log[\log(1/na_{2D}^2)]}{[\log(na_{2D}^2)]^2} \;,  
\label{BMF}
\end{equation}
which is expected to hold if $\log[\log(1/na_{2D}^2)]\gg 1$. For our smallest value of the gas parameter, $na_{2D}^2=10^{-7}$,
one finds $\log[\log(1/na_{2D}^2)]\simeq 2.8$. We fit the HD equation of state using the expression
\begin{eqnarray}
\frac{E-E_{MF}}{N} &=&\frac{2\pi\hbar^2n}{m} 
\label{FIT}
\\ \nonumber
&\times& \left[ -A\frac{\log[\log(1/na_{2D}^2)]}{[\log(na_{2D}^2)]^2}
+ \frac{B}{[\log(na_{2D}^2)]^2} \right]\;,  
\end{eqnarray}  
where $A$ and $B$ are free parameters. The best fit to all HD points, with the exclusion of the one at the highest density 
$na_{2D}^2=0.1$, gives $A=0.86$ and $B=2.26$ with $\chi^2/\nu=1.2$. The resulting curve is shown in Fig.~\ref{fig2} 
with a dashed line. We notice that the value $A=0.86$ is close to the prediction $A=1$ from the expansion (\ref{BMF}).  

In Fig.~\ref{fig3}, we show an enlargement of the equation of state in the gas parameter range $0.01\le na_{2D}^2\le 0.1$. 
In this 
region, the VMC results corresponding to the $V^{PP}(r)$ potential significantly deviate from the HD equation of state 
(thick dashed line). By using a polynomial fit to the PP equation of state (thick solid line), we calculate the inverse 
compressibility of the system $mc^2=n\partial\mu/\partial n$, where $c$ is the speed of sound and $\mu=dE/dN$ is the 
chemical potential. The inverse compressibility (thin solid line) exhibits a maximum and then drops to zero for 
$na_{2D}^2\simeq 0.04$. For comparison, the inverse compressibility of the HD gas as obtained from the fit (\ref{FIT}) is 
shown with a thin dashed line. We interpret the vanishing of $mc^2$ as the onset of instability against cluster formation.

The results for the condensate fraction corresponding to the HD and SD potentials are shown in Fig.~\ref{fig4} as a function
of the gas parameter. The condensate fraction, $N_0/N$, is obtained from the long-range behavior of the one-body density
matrix $N_0/N=\lim_{r\to\infty}\rho(r)$ (see Ref.~\cite{BC} for further details). The Bogoliubov theory applied to the 
2D Bose gas gives the result~\cite{Schick}
\begin{equation}
\frac{N_0}{N}=1+\frac{1}{\log(na_{2D}^2)} \;,
\label{CF}
\end{equation}
which should be contrasted with the 3D result $N_0/N=1-8\sqrt{na_{3D}^3}/(3\sqrt{\pi})$. From Fig.~\ref{fig4} one sees 
that for small values of the gas parameter the results of both the HD and SD potential agree with Eq. (\ref{CF}). For 
larger values of $na_{2D}^2$ the Bogoliubov result (\ref{CF}) overestimates the condensate fraction of the HD gas. In
the case of the SD potential we notice that $N_0/N$ first decreases by increasing the gas parameter and then increases.
This happens at densities where the interaction ranges of the particles start to overlap resulting in a reduction of 
interaction effects. It is worth noticing that in the region where Bogoliubov theory applies, for the same value of the 
gas parameter the condensate depletion is larger in 2D than in 3D, showing that quantum fluctuations are more 
effective in systems with reduced dimensionality.

Finally in Fig.~\ref{fig5} we report results of the pair distribution function $g_2(r)$ corresponding to the HD potential
for three different values of the gas parameter $na_{2D}^2=10^{-4}$, 0.01, 0.1. The function $g_2(r)$ gives the 
probability that two particles will be separated by a distance $r$. A shell-like structure in the pair distribution function 
is visible only for the highest density. The reduction from three to two dimensions changes the long-range behavior of $g_2(r)$ 
turning the dependence $(g_2^{3D}(r)-1)\propto 1/r^4$ into $(g_2^{2D}(r)-1)\propto 1/r^3$ and therefore the relevance of 
long-range correlations is enhanced in 2D.

\section{Conclusions}
\label{Conclusions}
We have carried out a study of the ground-state properties of a 2D homogeneous Bose gas with quantum Monte Carlo techniques. 
The universal behavior of the equation of state for small values of the gas parameter has been investigated by analyzing 
both mean-field and beyond mean-field contributions. By using a zero-range pseudopotential to describe interatomic interactions 
we have estimated that 
$na_{2D}^2\simeq 0.04$ is the critical density at which the system becomes unstable against cluster formation. These results are 
relevant for present and future experiments with ultracold Bose gases in highly anisotropic harmonic traps.

Acknowledgements: SP and SG acknowledge support by the Ministero dell'Istruzione, dell'Universit\`a e della 
Ricerca (MIUR). JB and JC acknowledge support from DGI (Spain) Grant No. BFM2002-00466 and Generalitat de 
Catalunya Grant No. 2001SGR-00222.

\end{document}